# Inelastic collisions and anisotropic aggregation of particles in a nematic collider driven by backflow

Oleg P. Pishnyak, Sergij V. Shiyanovskii, and Oleg D. Lavrentovich

Liquid Crystal Institute and Chemical Physics Interdisciplinary Program, Kent State University, Kent, OH 44242

We design a nematic collider for controlled out-of-equilibrium anisotropic aggregation of spherical colloidal particles. The nematic surrounding imparts dipolar interactions among the spheres. A bidirectional backflow of the nematic liquid crystal (NLC) in a periodic electric field forces the spheres to collide with each other. The inelastic collisions are of two types, head-to-tail and head-to-head. Head-to-tail collisions of dipoles result in longitudinal aggregation while head-to-head collisions promote aggregation in the transversal direction. The frequency of head-to-head collisions is set by the impact parameter that allows one to control the resulting shape of aggregates, their anisotropy and fractal dimension.

Particle collisions and ensuing aggregation represent a fascinating scientific theme. Typically, the particles interact through central forces that depend on distances but not on direction in space. Non-central interactions are of a greater challenge and promise in designing new materials [1], as demonstrated by studies of magnetic dipolar colloids [2-5], Janus particles [6-8], colloids with electric [9-13] and magnetic [14, 15] field-induced dipoles. Anisotropy of interactions is reflected in aggregation geometry. For example, diffusion limited aggregation of spheres with central forces produces structures with fractal dimension $d_f$ between 1 and 2 in two dimensions (2D) and between 1 and 3 in 3D, see, e.g., [16]. Once dipolar interactions are switched on, $d_f$ reduces dramatically, down to 1 in both 2D and 3D, reflecting head-to-tail chaining [2, 14, 15].

In this work, we demonstrate controlled out-of equilibrium aggregation of spherical particles with long-range anisotropic interactions induced by a NLC as a dispersive medium. Director distortions around the spheres are of dipole symmetry in the far-field limit. The elastic dipole $\mathbf{p} = (p_x, 0, 0)$ has two states, $p_x > 0$ and $p_x < 0$, separated by an energy barrier that prohibits their mutual transformations. The two types of particles, ">" and "<" are forced into two types of non-diffusive pair-wise collisions, "> >" (or "< <") and "$>\atop<$" in a specially designed collider with a bidirectional NLC flow driven by periodic electric pulses. "Particles ">" move unidirectionally in one plane, while particles "<" move in an antiparallel direction, located at an impact distance $b$ from the first plane. The inelastic collisions result in anisotropic aggregates with geometry controlled by $b$. By decreasing $b$ and enhancing the probability of the head-to-head collisions, one increases $d_f$ well above the value of 1 characteristic of dipolar systems with diffusion driven aggregation [2, 9-15].

Silica spheres of diameter $2R = (4.1 \pm 0.4)\,\mu m$ (Bangs Laboratories, Inc.) were dispersed in the NLC E7 (EM Industries) at (1-4) wt %. The particles were functionalized with *octadecyltrichlorosilane* (Sigma-Aldrich) to set perpendicular surface orientation of NLC. The cells were formed by glass plates with indium tin oxide (ITO) electrodes and rubbed polyimide PI2555 (Microsystems). Rubbing resulted in a uniform overall alignment described by a nonpolar director $\hat{\mathbf{n}}_0 \equiv -\hat{\mathbf{n}}_0$. It also produced a small ($\sim 1^0$) pretilt of $\hat{\mathbf{n}}_0$ that lifted the degeneracy of director rotation in the vertical electric field. The most effective control of impact parameter was observed for the cell thicknesses $6\,\mu m \leq d \leq 10\,\mu m$. The initial distribution of particles in the plane of the cell was random (with time, some particles tend to aggregate, but the aggregates can be re-dispersed by pressing on the top plate of the cell).

To match the uniform far-field $\hat{\mathbf{n}}_0 = \text{const}$ with the local radial director, each sphere produces a point defect, called a "hyperbolic" hedgehog [17], residing either on the left ($p_x > 0$, or ">" particle) or on the right side of the sphere ("<" particle), Fig.1a. The ensuing dipolar elastic interactions are used to assemble ordered colloidal arrays in the conditions close to equilibrium, for example, by moving the particles with optical tweezers [18-20]. Non-equilibrium assembly can be much richer than its equilibrium counterpart since the prehistory of particles' trajectories can be used as a control parameter. We design such a non-equilibrium assembly in an inelastic collider with a bidirectional "backflow", i.e., a bidirectional mass flow caused by director reorientation in the electric field.

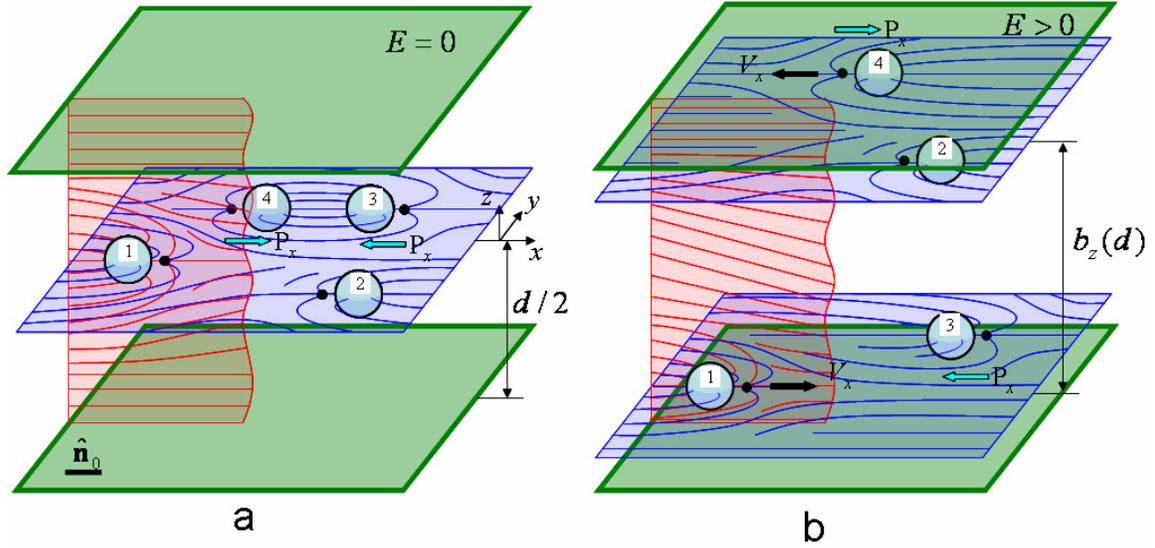

Fig. 1. Colloidal particles in a planar nematic cell. (a): particles levitate in the middle of cell in the absence of field; (b): the vertical electric field lifts the particles with oppositely oriented dipoles to two opposite bounding plates, where the backflow moves them in the two antiparallel horizontal directions.

In absence of the field, both ">" and "<" particles levitate near the middle of cell, $z = 0$, because of the elastic repulsion from the boundaries [21], Fig.1a. When an electric field is applied along the vertical $z$-axis, it causes three effects: (a) $\hat{\mathbf{n}}$ reorients towards the $z$-axis, adopting a symmetry of a letter "Z" with a maximum tilt at $z = 0$; (b) the ">" and "<" particles move to the opposite plates, fitting either the top or the bottom of Z-configuration of $\hat{\mathbf{n}}$ [21], Fig.1b; (c) director reorientation caused by periodic electric pulses leads to antisymmetric backflow along the horizontal axis $x$ with a non-zero period averaged displacement [21]. The backflow velocity is maximum near the top and the bottom plates and is zero at $z = 0$. As a result, the ">" and "<" particles are carried by two antiparallel flows, one close to the top and another close to the bottom. The impact parameter $b(d)$ controls the probability of "$\gtrless$" collisions but has little influence on "$>>$" and "$<<$" collisions, Fig.1b.

The cells were driven by rectangular pulses with RMS field $E \approx 0.5\ V/\mu m$, carrier frequency $f_c$=10 kHz (generator DS345, Stanford Research Systems), modulation frequency $f_m = 10\ Hz$ and duty ratio 50%. The dynamics was monitored with the Image-Pro Plus (Media Cybernetics) software. At least 3 experiments were performed for each sample's thickness; about 250 clusters were analyzed in each experiment.

The typical dynamics for $d = 10\ \mu m$ and $8\ \mu m$ is illustrated in Fig.2-4 (for video files see [22]). We used 20x and 5x objectives. In unpolarized light, the particles appear dark at a bright NLC background. The 20x objective resolved individual particles, while the 5x objective captured large areas. Comparing images of the same clusters, we find that the number of spheres resolved by the 20x objective is proportional to the number of black pixels resolved by the 5x objective with accuracy better than 20%. We analyse the structures using the 'inertia' tensor defined for a cluster of $N$ black pixels as:

$$M_{\alpha\beta} = \sum_{i=1}^{N} \left(\rho_\alpha^i - \bar{\rho}_\alpha\right)\left(\rho_\beta^i - \bar{\rho}_\beta\right), \qquad (1)$$

where $\alpha, \beta = x, y$, $\boldsymbol{\rho}^i = \{\rho_x^i, \rho_y^i\}$ are coordinates of the $i$-th black pixel and $\bar{\boldsymbol{\rho}} = N^{-1}\sum_{i=1}^{N}\boldsymbol{\rho}^i$ is a center of mass of the cluster. The inertia tensor allows us to define a characteristic diameter $D = 2\sqrt{\mathrm{Tr}\mathbf{M}/N}$ and anisotropy $q$ of the cluster:

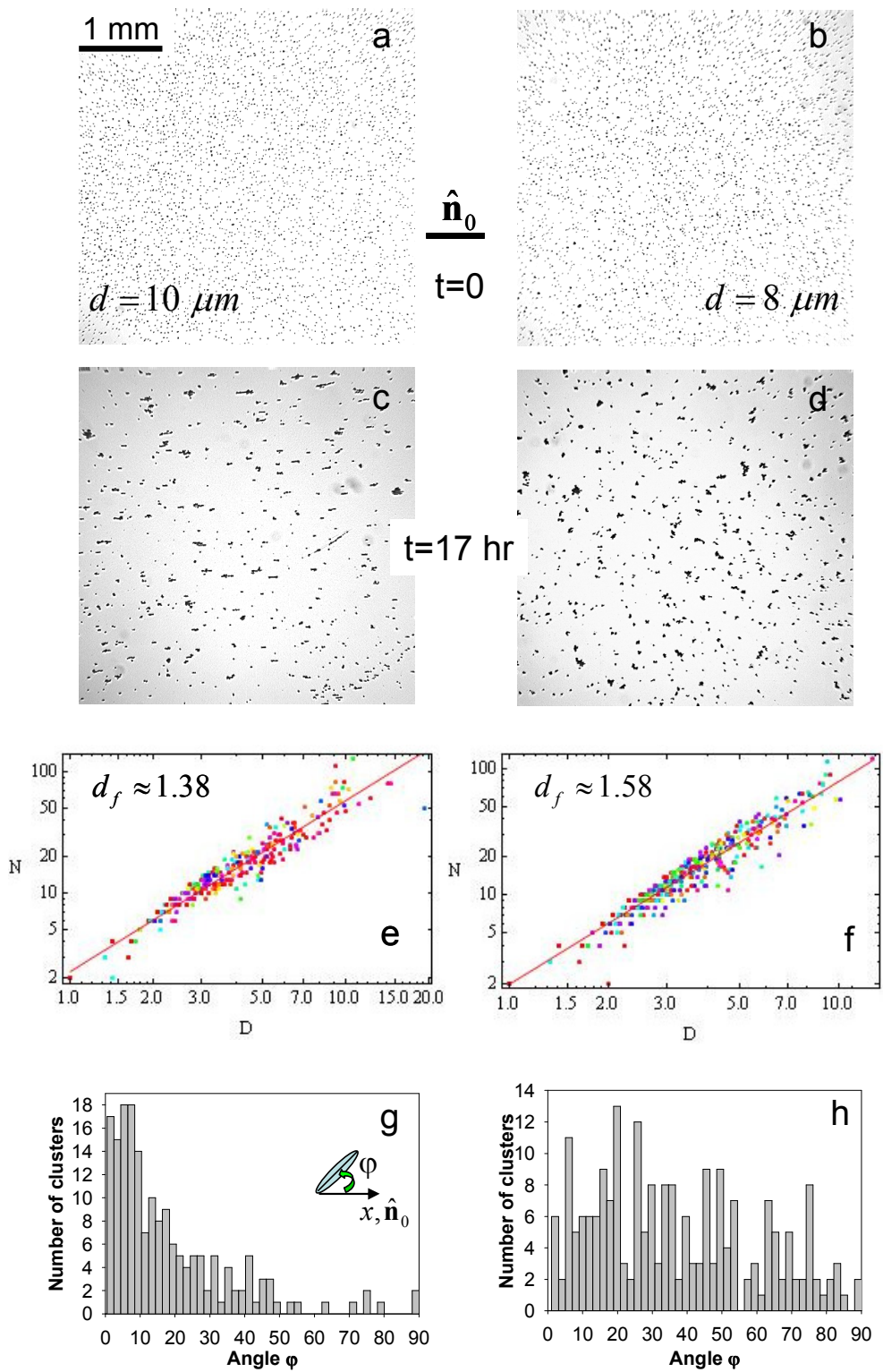

Fig. 2. Aggregation dynamics of particles in $10\,\mu m$ (a,c,e,g) and $8\,\mu m$ (b,d,f,h) cells at the beginning (a, b, $t=0$) and at the end (c,d, $t=17$ hours) of the modulated electric field

signal. (e,f): log-log plot of the number of pixels $N$ vs cluster diameter $D$ fitted with $\log N = d_f \log D + const$ after $t=17$ hours; (g,h): orientational distribution of clusters after $t=17$ hours; $\varphi$ is the angle between the long axis of a cluster and $\hat{\mathbf{n}}_0$. The cells are filled with the mixture E7/silica particles (3 wt. %).

$$q = (M_{xx} - M_{yy})/(M_{yy} + M_{xx}); \qquad (2)$$

$q = 0$ for a disk-like cluster, $q = 1$ for a line cluster along the $x$-axis (parallel to $\hat{\mathbf{n}}_0$) and $q = -1$ for a line cluster along the $y$-axis. The relationship between $N$ and $D$ is close to fractal, $N \propto D^{d_f}$, Fig.2e,f.

There are two opposite tendencies in cluster formation, namely, chaining in the direction of $\hat{\mathbf{n}}_0$ (Fig.4a) and perpendicular to it (Fig.4b); they result in a number of intermediate geometries, Fig.4c and, over a sufficiently long time, in jammed structures, Fig.4d. The two opposite tendencies result in different shape of clusters in thick and thin cells: $d_f$ is smaller and $q$ is larger in $d = 10\ \mu m$ cells as compared to $d = 8\ \mu m$ cells, Fig.2,3. When $d \gg 2R$ and the concentration of particles is low, then only linear chains of parallel dipoles form along $\hat{\mathbf{n}}_0$, see Fig.2 in Ref. [21].

The kinetics and geometry of cluster formation can be qualitatively explained by considering the balance of forces: (a) elastic attractive forces and (b) forces that resist aggregation and keep the ">" and "<" particles close to the opposite boundaries.

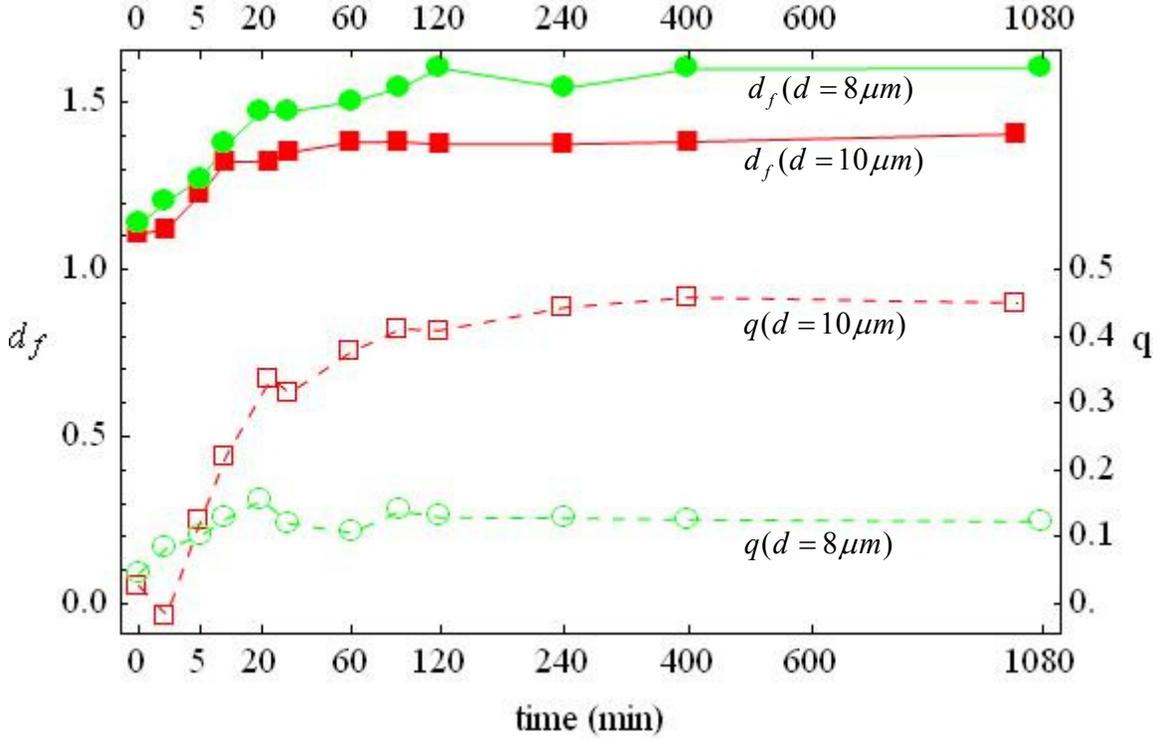

Fig. 3. Time dependence of fractal dimension $d_f$ (filled markers) and anisotropy $q$ (empty markers) of formed clusters in 10 $\mu m$ (squares) and 8 $\mu m$ (circles) cells.

(a) The interaction potential of two elastic dipoles located at $\mathbf{r}^i$ and $\mathbf{r}^j$ [17] is

$$U_{pp} = 12\pi K \left[ \frac{\mathbf{p}^i \mathbf{p}^j}{r^3} - 3\frac{(\mathbf{p}^i \mathbf{r})(\mathbf{p}^j \mathbf{r})}{r^5} \right],$$ where $r = |\mathbf{r}| \gg 2R$ and $\mathbf{r} = \mathbf{r}^i - \mathbf{r}^j$, $K$ is the average elastic constant of the NLC, $K \approx 15\, pN$ for E7; $\mathbf{p}^i = (p_x^i, 0, 0)$ is the dipole of the i$^{th}$ particle, directed from the hedgehog towards the sphere; $p_x = \pm 2.04 R^2$; we neglect higher miltipole corrections. There are two extreme collision scenarios. One is the head-to-tail chaining, $p_x^i p_x^j > 0$, along $\hat{\mathbf{n}}_0$, Fig.4a, a well-studied case of equilibrium

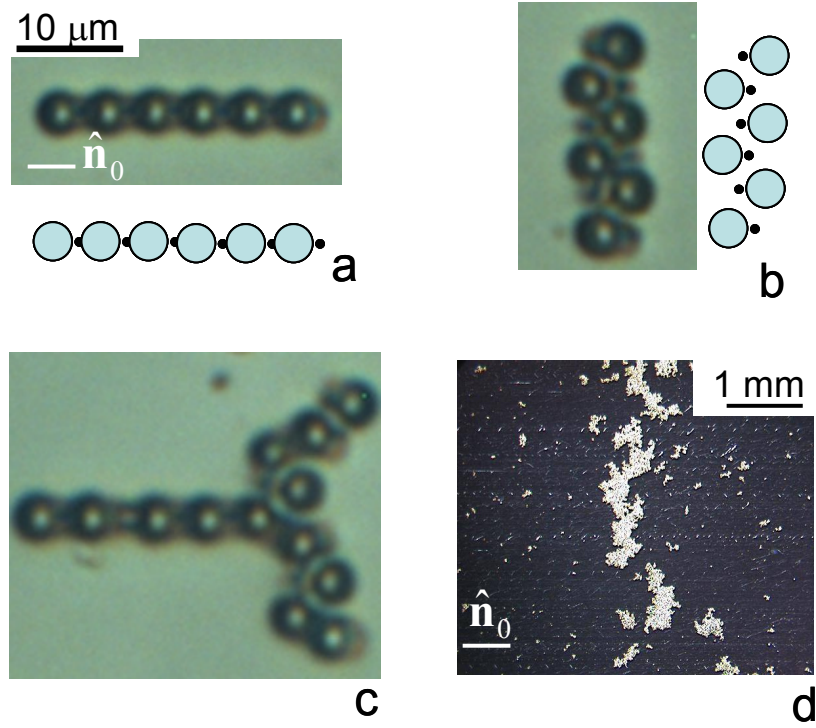

Fig. 4. Different geometries of clustering (a): linear chaining parallel to the director $\hat{\mathbf{n}}_0$ by particles with the same polarity of the elastic dipoles; (b): zigzag chaining perpendicular to $\hat{\mathbf{n}}_0$ by particles with alternating polarity; (c): aggregate growing in both directions, perpendicular and parallel to $\hat{\mathbf{n}}_0$; (d): colloidal particles formed an elongated cluster jamming the collider in the direction perpendicular to backflow after application of a modulated electrical signal with $E \approx 0.5\ V/\mu m$, $f_c = 10$ kHz, $f_m = 10$ Hz; $d \approx 9\ \mu m$. The cell was filled with the mixture E7/silica particles (3 wt. %).

aggregation [17]. The second is arrangement of antiparallel dipoles, $p_x^i p_x^j < 0$, in a zigzag perpendicular to $\hat{\mathbf{n}}_0$, Fig.4b [18]. In the first case, $d_f \to 1$ and $q \to 1$; in the second case, $d_f \to 1$ and $q \to -1$. The impact parameter $b(d)$ controls the second type of collisions, as the first occurs among the particles located in the same backflow stream. In its turn, $b(d)$ depends on the elastic and dielectrophoretic forces discussed below.

(b1) The electric field causes director gradients near the boundaries over the length $\xi \approx \frac{1}{E}\sqrt{\frac{K}{\varepsilon_0 \Delta\varepsilon}}$ [16], $\varepsilon_0 \approx 8.85 \cdot 10^{-12}$ $C^2/N \cdot m^2$ is the permittivity of free space; $\Delta\varepsilon = \varepsilon_\parallel - \varepsilon_\perp = 19.0 - 5.2 = 13.8$ is the dielectric anisotropy of E7. The sphere moves from the uniformly titled center of the cell to the boundary to replace the distorted region and to reduce the elastic energy, by $\Delta f_{trap} \sim KR^3/\xi^2$ for $R \leq \xi$ [23] and by $\Delta f_{trap} \sim KR$ for $R \geq \xi$. In a strong field ($E = 0.5$ $V/\mu m$, $\xi \approx 0.7$ $\mu m$), the elastic trapping force for the particle near the substrate is $F_{trap} \sim KR/\xi \approx 40$ pN.

(b2) The second force $F_D$ keeping the particles near the boundaries is caused by the dielectric permittivity gradients. As one moves from $z = 0$ to the boundaries, the effective dielectric permittivity changes from $\varepsilon(0) \approx \varepsilon_\parallel$ to $\approx \varepsilon_\perp$. The particles move to the region with the lowest $\varepsilon$, to minimize the dielectric energy. This effect is similar to the dielectrophoretic effect in isotropic media [24], with an important difference. In the isotropic medium, the dielectrophoretic force $\sim \nabla |\mathbf{E}|^2$ is created by a special geometry of electrodes. In the anisotropic and nonuniform medium, the dielectrophoretic force occurs even if the electrodes are flat and parallel, as long as $\varepsilon$ varies in space [25]. In our case, $\varepsilon(z) = \varepsilon_\perp \sin^2 \theta(z) + \varepsilon_\parallel \cos^2 \theta(z)$, where $\theta(z)$ is the angle between $\mathbf{E}$ and $\hat{\mathbf{n}}$. To estimate this dielectrophoretic force, we neglect the in-plane components of the field. The dielectric contribution $f_D$ to the free energy that depends on the particle coordinate $z_p$, writes as $f_D = -\pi\varepsilon_0 g U^2 \int_0^R \left\{ \left[ d\varepsilon_{eff}^{-1} + \int_{-z_r}^{z_r} \left( \varepsilon_p^{-1} - \varepsilon^{-1}(z_p+z) \right) dz \right]^{-1} - \varepsilon_{eff}/d \right\} r\, dr$, where

$\varepsilon_{eff} = d / \int_{-d/2}^{d/2} \varepsilon^{-1}(z) dz$, $z_r = (R^2 - r^2)^{1/2}$, $\varepsilon_p$ is the dielectric constant of particle, $g$ is a phenomenological factor that accounts for the contribution of the point defect, and $U$ is the applied voltage. With $\varepsilon_p = 3.8$ [26], $g = 1.5$, $E = U/d = 0.5\ V/\mu m$, the dielectrophoretic force is $F_D \approx 40$ pN for both $d = 8\ \mu m$ and $d = 10\ \mu m$. The behavior of $F_D$ is similar to that of $F_{trap}$: $F_D \approx 0$ near $z = 0$ where the gradient of $\varepsilon$ vanishes, but increases towards the bounding plates. The sum $F_{trap} + F_D \approx 80$ pN is larger than the gravity forces ($F_g \approx 0.3$ pN) and elastic repulsive forces ($F_{wall} \approx 40$ pN) between the particles and bounding walls [21].

Under the modulated signal, the particles move into opposite directions parallel to the rubbing direction ($x$-axis) in accordance with the generated antisymmetric backflow [21]. The attractive force between ">" and "<" particles separated by a distance $r_z$ along the $z$-axis is $F_{><} = -\nabla U_{><} \sim 150\pi K R^4/b^4$ [18], where $b = (b_y^2 + b_z^2)^{1/2}$, $b_y$ and $b_z$ are horizontal and vertical distances between the particles. For the thick cells, $d = 10\ \mu m$, $r_z \approx 6\ \mu m$ in a strong field $E = 0.5\ V/\mu m$, the maximum $(b_y = 0)$ inter-particle force is $F_{><} \approx 100\ pN$, which is only slightly larger than the sum $F_{trap} + F_D$. As a result, ">" particles slip past "<" particles without coalescence. Particles moving in the same direction can still aggregate along $\hat{\mathbf{n}}_0$ through the scenario ">>" or "<<", Fig.4a. The resulting $d_f$ is low and the shape anisotropy is relatively high, even at late stages, Fig.3.

In thin cells, $d = 8$ $\mu m$, $r_z \approx 4$ $\mu m$, the attractive forces (a) are more significant as compared to (b1, b2) forces, $F_{><} \approx 500$ $pN$, facilitating aggregation perpendicular to $\hat{\mathbf{n}}_0$, Fig.4b. The characteristic velocity of aggregation $v_{agg} \approx F_{><}/(6\pi R \alpha_4) \approx 50$ $\mu$m/s, where $\alpha_4 = 0.225$ Pa·s is the Leslie coefficient [27], is much larger than an average velocity of particles $v \approx 2$ $\mu$m/s [21]. This transversal chaining, together with the chaining along $\hat{\mathbf{n}}_0$, produces clusters of higher $d_f$ and smaller $q$, Fig.2,3.

We note that the dipolar model used for analytical estimates is only a rough approximation. Its validity is diminished by the following factors: (a) shortness of separating distances at the stage of collisions; (b) field-induced torques that modify the interactions, see the review [28]; (c) high velocities, as the corresponding Ericksen number (the ratio of elastic and viscous torques) $Er = \alpha_4 v_{agg} R / K$ becomes larger than 1, which means that $\hat{\mathbf{n}}$ is modified by flows [29-31]. However, the outcome of collisions, the geometry of clusters, is dictated primarily by the relative number of ">" and "<" particles in a given cluster.

We demonstrated a new approach to form anisometric clusters through non-equilibrium collisions in a nematic collider with two anti-parallel flows of ">" and "<" particles. The clustering geometry can be controlled by the cell thickness, amplitude of the electric field, size and concentration of particles. The proposed colloidal collider with bidirectional flow presents a rich experimental system for further studies of anisotropic clustering and jamming.

We acknowledge support NSF DMR 0906751 and DOE grant DE-FG02-06ER 46331.